\newcommand{\beq}{\begin{equation}}
\newcommand{\eeq}{\end{equation}}
\newcommand{\beqa}{\begin{eqnarray}}
\newcommand{\eeqa}{\end{eqnarray}}
\newcommand{\ket}[1]{| #1 \rangle}
\newcommand{\bra}[1]{\langle #1 |}

\newcommand{\hone}{\hat{h}_1}
\newcommand{\htwo}{\hat{h}_2}

\documentclass[prl,twocolumn,showpacs]{revtex4}

\usepackage{epsf}

\begin{document}

\title{Two-photon imaging and quantum holography}

\author{Gunnar Bj\"{o}rk}
\email{gunnarb@imit.kth.se} \homepage{http://www.ele.kth.se/QEO/} 
\affiliation{Department of Microelectronics and Information 
Technology, Royal Institute of Technology (KTH), Electrum 229, 
SE-164 40 Kista, Sweden}

\author{Jonas S\"{o}derholm}
\affiliation{Department of Microelectronics and Information 
Technology, Royal Institute of Technology (KTH), Electrum 229, 
SE-164 40 Kista, Sweden}

\author{Luis L. S\'{a}nchez-Soto}
\affiliation{Departamento de \'{O}ptica, Facultad de Ciencias 
F\'{\i}sicas, Universidad Complutense, 28040 Madrid, Spain}

\date{\today}

\begin{abstract}
It has been claimed that ``the use of entangled photons in an 
imaging system  can exhibit effects that cannot be mimicked by 
any other two-photon source, whatever strength of the 
correlations between the two photons'' [A. F. Abouraddy, B. E. A. 
Saleh, A. V. Sergienko, and M. C. Teich, Phys. Rev. Lett. 
\textbf{87}, 123602 (2001)]. While we believe that the cited 
statement is true, we show that the method proposed in that 
paper, with ``bucket detection'' of one of the photons, will give 
identical results for entangled states as for appropriately 
prepared classically correlated states.
\end{abstract}


\maketitle

Recently, there has been a constant focus of  interest in quantum 
limited imaging~\cite{Kolobov,Kolobov 2,Fabre} and 
lithography~\cite{Boto,Kok,Bjork,Bjork 2}. In the former of the 
applications, the quantum correlation between the amplitude 
fluctuations of an appropriately entangled multimode state of 
light is used to beat the standard quantum limit of spatial 
resolution. The latter application capitalizes on the short de 
Broglie wavelength of entangled multiphoton states and offers, in 
principle, a spatial resolution that is independent of the 
classical wavelength of the electromagnetic field~\cite{Jacobson}.

In a recent paper by Abouraddy \textit{et al.}~\cite{Abouraddy}, 
a method has been proposed where an entangled photon pair is used 
to image an object. The setup is depicted in Fig.~\ref{Fig: 
Setup}. Each of the photons in a photon pair is sent through an 
object. The objects labeled 1 and 2 are described by the impulse 
response operators $\hone$ and $\htwo$, respectively. The claim 
by Abouraddy \textit{et al.} is that, if the source emits an 
entangled two-photon multimode state, and if one of the photons 
emitted is detected by a ``bucket detector'' (to be defined 
below), information about a test object is obtained that cannot 
be obtained with any classical source of light. Abouraddy 
\textit{et al.} have proposed two schemes, in the first, called 
distributed quantum imaging, the test object, that we want to 
characterize, is described by $\hone$, while $\htwo$ is a (known) 
reference object \cite{Abouraddy}. The authors have also 
suggested a scheme where the roles of the objects 1 and 2 are 
reversed ~\cite{Abouraddy 2}, so that the test object is 
described by $\htwo$ and the photon that goes through it is 
detected by a bucket detector, while the photon going through the 
reference object, now described by $\hone$, is detected by a 
detector array. This imaging method was named quantum holography.

The claim by Abouraddy \textit{et al.} has been put in doubt by a 
recent analysis, and an experiment, by Bennink \textit{et 
al.}~\cite{Bennink}. Their analysis showed that if the test 
object in a quantum holographic setup is lossless, no information 
about the test object is obtained. In their experiment they used 
a lossy test object and a correlated classical source of light. 
Yet, using what Bennink \textit{et al.} called ``'two-photon' 
coincidence imaging'' (where the authors' quotation marks 
indicate that the detection method mimics what Abouraddy 
\textit{et al.} call quantum holography, but that classically 
correlated states were used instead of two-photon states) the 
information stored in the test object could be read out. In this 
paper, we extend the results of Bennink \textit{et al.} by 
showing that distributed quantum imaging with a lossless 
reference object gives identically the same information about the 
test object as if an appropriate classical source were used. We 
also show that quantum holography with a lossless reference 
object (as in the experiment by Bennink \textit{et al.}) can 
always be mimicked by a classical source.

\begin{figure}
\epsfxsize=8cm
\epsfbox{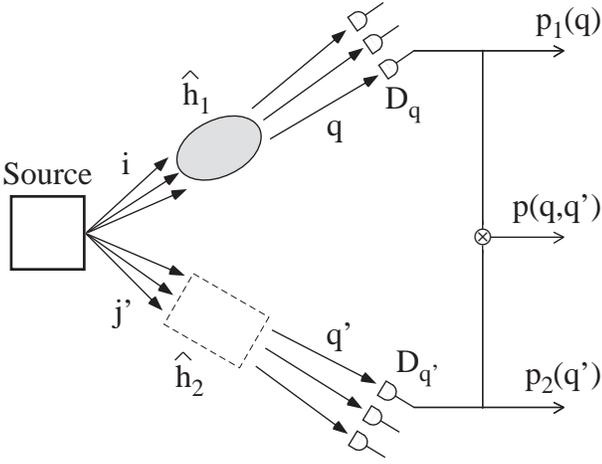}
\caption{The two-photon source emits one photon into the modes 
denoted $i$, which propagate through object 1 described by the 
transformation $\hone$. The linear impulse response operator 
$\hone$ contains the information about object 1. Another photon 
is emitted into modes $j'$ and is subjected to the operator 
$\htwo$ describing object 2. On the output side, there is an 
array of detectors, each measuring the photon number of a 
specific output mode.} \label{Fig: Setup}
\end{figure}

We shall assume that the considered modes, both on the source 
side and on the detection side, are discrete. Apart from this 
difference, we shall try to keep the notation as similar as 
possible to the notation in Ref.~\cite{Abouraddy}. Moreover, we 
will, just as Abouraddy \textit{et al.}~\cite{Abouraddy}, assume 
that the two photons are spatially filtered in such a way that 
the photon emitted into unprimed modes does not interact with 
object 2, and the photon emitted into primed modes does not 
interact with object 1.

Consider a source that emits light in the pure two-photon state 
\beq \ket{\Psi} = \sum_i \sum_{j'} \varphi(i,j') \ket{1_i,1_{j'}} 
, \label{Eq: initial state} \eeq where we have used the shorthand 
notation $\ket{1_i}\otimes \ket{1_{j'}} \equiv \ket{1_i,1_{j'}}$. 
Let us first see what happens if we completely ignore the photons 
passing through object 2. Writing $\hat{\rho} = 
\ket{\Psi}\bra{\Psi}$, the state of the unprimed photon then 
becomes \beq \textrm{Tr}_2 ( \hat{\rho} ) = \sum_{q'} 
\bra{1_{q'}} \hat{\rho} \ket{1_{q'}} . \label{Eq: density 
operator mixed state} \eeq In general, this is a mixed state of a 
single photon. After interaction with object 1, the output state 
becomes \beq \hone \sum_{q'} \bra{1_{q'}} \hat{\rho} \ket{1_{q'}} 
\hone^\dagger =  \sum_{q'} \bra{1_{q'}} \hone \hat{\rho} 
\hone^\dagger \ket{1_{q'}} , \label{Eq: single photon output 
state} \eeq where the equality follows from the fact that we have 
assumed that $\hone$ does not operate on the primed vectors. It 
may be prudent, at this point, to establish that the mode sets on 
the input and output side of object 1 or 2 do not have to be of 
the same type, as long as both are enumerable and orthonormal. 
For any given object, every specific pair of mode sets will lead 
to a unique transformation $\hone$. On the detector side it is 
convenient to work in the eigenmodes of the detectors. Then, the 
probability $p(q)$ of a photon detection in mode (detector) $q$ is
\begin{equation}
p_1(q)  = \sum_{q'} \bra{1_q,1_{q'}} \hone \hat{\rho} 
\hone^\dagger \ket{1_q,1_{q'}}. \label{Eq: single photon 
probability}
\end{equation}
If one defines $\gamma_1(i,j) = \sum_{k'} \varphi(i,k') 
\varphi^\ast(j,k')$, the coefficients $\gamma_1(i,j)$ turn out to 
be the density operator coefficients $\bra{1_i}\textrm{Tr}_2 ( 
\hat{\rho} ) \ket{1_j}$, so that \beq \textrm{Tr}_2 ( \hat{\rho} 
) = \sum_i \sum_j \gamma_1(i,j) \ket{1_i} \bra{1_j} . \label{eq: 
density operator expression} \eeq By inserting the resolution of 
the identity twice into Eq.~(\ref{Eq: single photon 
probability}), one arrives at \beq p_1(q)=\sum_i \sum_j 
\gamma_1(i,j) h_1(q,i) h_1^\ast(q,j) , \label{Eq: final 
probability expression} \eeq where $h_1(q,i) \equiv \bra{1_q} 
\hone \ket{1_i}$ and $h_1^\ast(q,j) \equiv (\bra{1_q} \hone 
\ket{1_j})^\dagger$. The expression above is the direct 
discrete-mode equivalent to Eq.~(8) in Ref.~\cite{Abouraddy}. If 
one permutes the indices $q$ and $q'$, and lets the index $1 
\rightarrow 2$ in Eqs.~(\ref{Eq: single photon probability}) and 
(\ref{Eq: final probability expression}), one obtains the 
expressions for $p_2(q')$.

\begin{figure}
\epsfxsize=8cm
\epsfbox{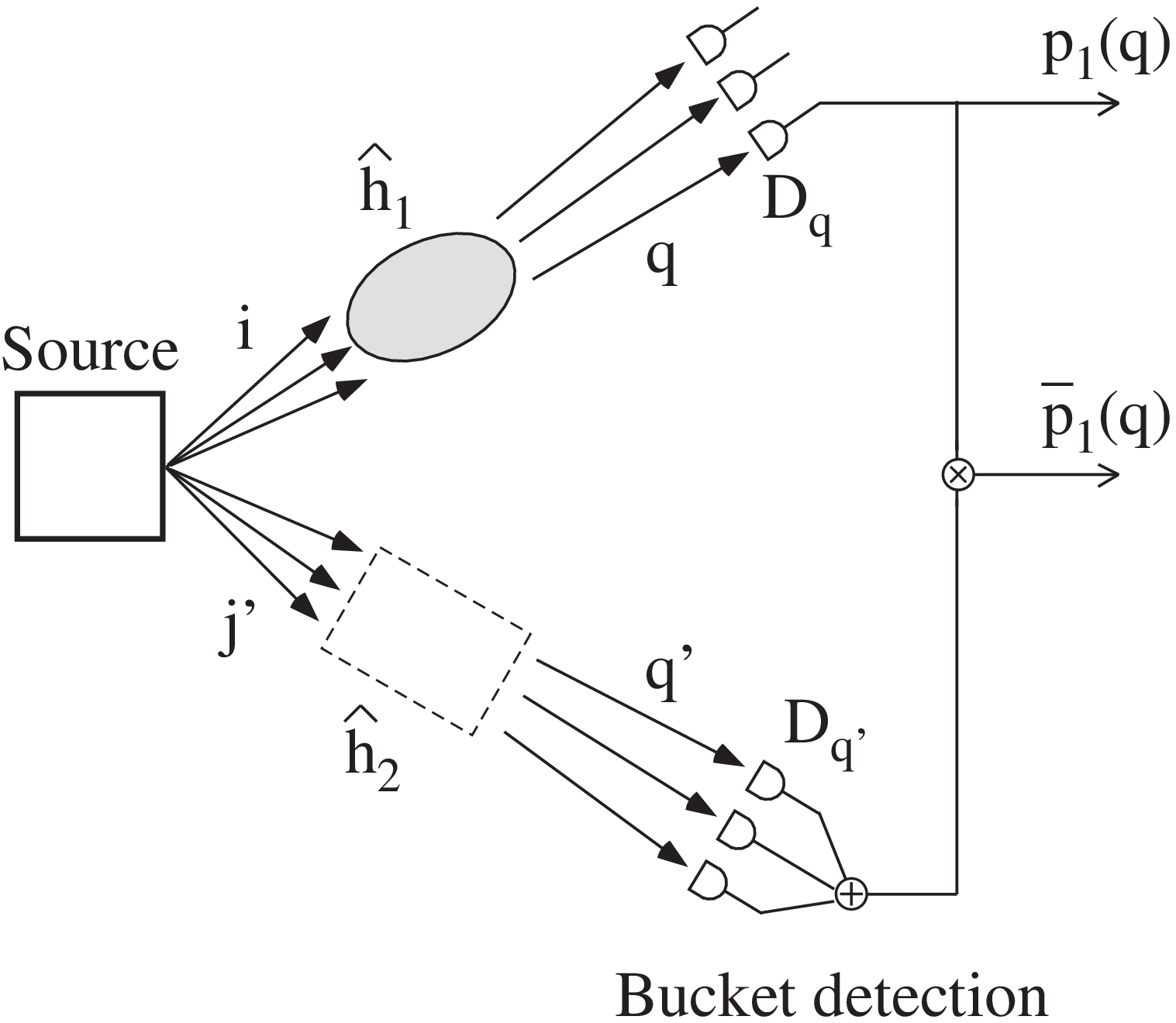}
\caption{Schematic drawing of ``bucket detection'' of the photon 
emitted into the primed modes corresponding to the lower detector 
array in the figure.} \label{Fig: bucket detector}
\end{figure}

Now, assume that both photons in the photon pair are considered. 
The output state after the interaction with the two objects 
becomes $\hone \htwo \hat{\rho} \htwo^\dagger \hone^\dagger$. 
Note that, since $\hone$ and $\htwo$ operate in different vector 
spaces, they commute. The probability of registering a correlated 
detection event between detector $q$ and detector $q'$ can be 
expressed as \beq p(q,q') =  \bra{1_q,1_{q'}} \hone \htwo 
\hat{\rho} \htwo^\dagger \hone^\dagger \ket{1_q,1_{q'}} . 
\label{Eq: two photon probability} \eeq

A ``bucket detector'' is a multimode detector where all the modes 
propagating through an object are measured jointly. Hence, the 
information about the location of the detected photon (or 
equivalently, in what mode the photon was detected) is 
``erased.'' A schematic of such a detector is depicted in 
Fig.~\ref{Fig: bucket detector}. In this case, the probability of 
photon detection in mode $q$ (the marginal probability 
distribution of the photons interacting with object 1) becomes 
\beq \bar{p}_1(q) =  \sum_{q'} \bra{1_q,1_{q'}} \hone \htwo 
\hat{\rho} \htwo^\dagger \hone^\dagger \ket{1_q,1_{q'}} . 
\label{Eq: bucket probability} \eeq If we assume that the state 
from the source is entangled such that $ \varphi(i, j) = 
\varphi(i) \delta_{ij'}$, Eq.~(\ref{Eq: bucket probability}) can 
be rewritten as \beq \bar{p}_1(q) = \sum_i \sum_j \sum_{q'} 
\varphi(i) \varphi^\ast(j) h_1(q,i) h_1^\ast(q,j) h_2(q',i) 
h_2^\ast(q',j) . \label{Eq: Expanded probability} \eeq Following 
Ref.~\cite{Abouraddy}, we also introduce the parameter 
$g_2(k',l') = \sum_{q'} h_2(q',k') h_2^\ast(q',l')$. Replacing 
the sum over $q'$ in Eq.~(\ref{Eq: Expanded probability}) with 
this expression, we can write \beq \bar{p}_1(q) = \sum_i \sum_j  
\varphi(i) \varphi^\ast(j) g_2(i,j) h_1(q,i) h_1^\ast(q,j) . 
\label{Eq: Expanded probability 2} \eeq This is the discrete 
counterpart of Eq.~(12c) in Ref.~\cite{Abouraddy}.  Comparing 
expressions (\ref{Eq: final probability expression}) and 
(\ref{Eq: bucket probability}), Abouraddy \textit{et al.} claim 
that ``based on classical probability analysis one would 
intuitively expect that $p_1(q)$ would be equal to 
$\bar{p}_1(q)$. This is not always the case, however.'' From a 
strictly mathematical point of view it is evident that $p_1(q)$ 
and $\bar{p}_1(q)$ are not equal for general functions $h_1(q,i)$ 
and $h_2(q',i')$. However, we shall now prove that if the 
reference object is lossless, the two expressions are equal.

Assume, therefore, that object 2, described by $\htwo$ is 
lossless. This immediately implies that \textit{the operator 
$\htwo$ is unitary}. Using this fact, we can write the expression 
for $p_1(q)$ as \beq p_1(q) = \sum_{q'} \bra{1_q,1_{q'}} \hone 
\htwo^\dagger \htwo \hat{\rho} \htwo^\dagger \htwo \hone^\dagger 
\ket{1_q,1_{q'}}. \label{Eq: using unitarity} \eeq Furthermore, 
since $\hone$ and $\htwo$ commute, we can recast this equation as 
\beq p_1(q) =  \sum_{q'} \bra{1_q,1_{q'}} \htwo^\dagger \hone 
\htwo \hat{\rho} \htwo^\dagger \hone^\dagger \htwo 
\ket{1_q,1_{q'}} . \label{Eq: using unitarity 2} \eeq Since the 
trace operation is basis invariant, the trace Tr$_2$ of any 
operator $\hat{O}$ can be expressed either 
$\sum_{q'}\sum_{n=0}^\infty \bra{n_{q'}}\hat{O}\ket{n_{q'}}$ or 
$\sum_{q'} \sum_{n=0}^\infty \bra{n_{q'}} \htwo^\dagger \hat{O} 
\htwo \ket{n_{q'}}$, where $\htwo$ is an arbitrary, unitary 
operator. Hence, from Eq. (\ref{Eq: using unitarity 2}), we get 
\beq p_1(q) =  \sum_{q'} \bra{1_q,1_{q'}} \hone \htwo \hat{\rho} 
\htwo^\dagger \hone^\dagger \ket{1_q,1_{q'}} = \bar{p}_1(q). 
\label{Eq: using unitarity 3} \eeq This relation implies that 
Eqs. (\ref{Eq: single photon probability}) and (\ref{Eq: bucket 
probability}) are \textit{identical} if the object 2 (the 
reference object) is lossless, irrespective of $\hat{\rho}$ and 
$\hone$. Hence, in this case, nothing is gained by using 
distributed quantum imaging over single photon (per necessity, 
uncorrelated) imaging. The single-photon state $\textrm{Tr}_2 ( 
\hat{\rho} )$ will give the same detection statistics as $ 
\hat{\rho}$.

Let us now consider the case where the test object is described 
by $\htwo$, that is, it is the modes through the test object that 
are detected with the ``bucket detector.'' This imaging principle 
is called ``quantum holography''~\cite{Abouraddy 2} or 
``two-photon coincidence imaging'' ~\cite{Bennink}. Abouraddy 
\textit{et al.}~\cite{Abouraddy 2} claimed that it is possible to 
use entanglement between the states in the primed and unprimed 
modes to read out holographic information about the test object 
$\htwo$ even if the photons traversing it are detected with a 
bucket detector, such as a sphere coated on the inside with a 
photosensitive film that surrounds it. However, Bennink 
\textit{et al.} showed that if the test object is lossless, then 
the bucket detector always clicks, effectively tracing out all 
information about the test object. Hence, quantum holography of 
such test objects does not work.

However, assume that the test object (object 2) is lossy. It is 
clear that in this case the reference object (object 1) should be 
lossless, as losses will only introduce additional randomness in 
the measurement. Since $\hone$ and $\htwo$ commute, this 
randomness corresponds to loss of information about $\htwo$. If 
the reference object is lossless, it is rather obvious that if a 
two-photon (possibly) entangled state $\ket{\Psi}$ is used, its 
quantum holography detection statistics will be exactly mimicked 
by the classically correlated state \beq \hat{\rho}' = \sum_i 
\hone^\dagger\ket{1_i}\bra{1_i} \hone \otimes \bra{1_i} \hone 
\hat{\rho} \hone^\dagger \ket{1_i} . \label{eq:classical state} 
\eeq Hence, it seems to us that entangled-state holography with 
bucket detection \textit{of any object} is of limited use, since 
identical results can be obtained using classical states. This 
was what Bennink \textit{et al.}~\cite{Bennink} demonstrated 
experimentally. The statistical distribution $\bar{P}_1(q)$ in 
their experiment very clearly brought out the information encoded 
in the mask although all the photons interacting with the (lossy) 
mask were detected by a bucket detector.

Above, we have shown that distributed quantum imaging and quantum 
holography with a lossless reference object (but, in general, a 
lossy test object) will not offer anything imaging with classical 
states cannot provide. Bennink \textit{et al.}~\cite{Bennink} 
proved that quantum holography of a lossless test object also can 
be mimicked by classical states. Now, we will treat the case when 
both object 1 and 2 are lossy. The standard way of including 
losses is to extend the two original sets of modes with auxiliary 
modes. We shall do so by assuming that the respective detector 
eigenmodes constitute only a subset of the primed and unprimed 
modes. To be realistic we can assume that only the photons in $N$ 
($N'$) of the unprimed (primed) modes are detected, and that 
these modes are labeled $1,2, \ldots, N$ ($1',2', \ldots, N'$). 
The probability of detecting a photon in mode $q$ then becomes
\begin{eqnarray}
P_1(q) & = & \sum_{q'=1}^{N'}  \bra{1_q,1_{q'}} \hone \htwo 
\hat{\rho} \htwo^\dagger \hone^\dagger
\ket{1_q1_{q'}} \nonumber \\
& & +  \sum_{q'> N'} \bra{1_q,1_{q'}} \hone \htwo \hat{\rho} 
\htwo^\dagger \hone^\dagger
\ket{1_q,1_{q'}} \nonumber \\
& & = \bar{P}_1(q) + P_{1,0}(q). \label{Eq: single photon 
probability with loss}
\end{eqnarray}
From Eq.~(\ref{Eq: single photon probability with loss}), we see 
that photodetection probability $P_1(q)$ is simply the sum over 
all the detection events $q'$ of the two-photon joint probability 
distribution $\bar{P}_1(q)= \sum_{q'=1}^{N'} P(q,q')$ and the 
probability of detecting one photon in mode $q$ and no photon in 
the modes $q'=1',2', \ldots,N'$. This is just what one expects 
intuitively. For lossy reference objects, $P_1(q)$ differs from 
$\bar{P}_1(q)$.

Now, denote the probability of not detecting the photon in the 
primed modes $P_0=\sum_q P_{1,0}(q)$ and consider,  e.g., the 
product state \beq \sum_{j'=1'}^{N'}\bra{1_{j'}}\htwo \hat{\rho} 
\htwo^\dagger \ket{1_{j'}} \otimes \htwo^\dagger\left ( 
\ket{1_{1'}}\bra{1_{1'}} + {P_0 \ket{1_{M'}}\bra{1_{M'}} \over 
1-P_0} \right ) \htwo , \label{eq: example state} \eeq where $M' 
> N'$. It is quite obvious that this state, that lacks any 
correlation between the photon in the primed and unprimed modes, 
will yield an identical photodetection statistics distribution 
$\bar{P}_1(q)$ to the corresponding distribution of the state 
$\hat{\rho}$ (that may be entangled). Note that this is true 
irrespective of $\hat{\rho}$, $\hone$, and $\htwo$. The result 
depends critically on the fact that $\htwo$ is an operation local 
in the primed modes. However, this is, in general, not a 
physically accessible state since it may contain excitation in 
modes with $M'>N'$ (the modes corresponding to the loss). If only 
the modes $q'=1', \ldots , N'$ can be excited, then one cannot, in 
general, mimic the photodetection statistics of the two lossy 
objects illuminated by an two-photon entangled state, with a 
classically correlated state. However, even if this is not the 
case, we do not see what could be gained by using entanglement, 
because, in general, loss will destroy the entanglement of a 
quantum state more rapidly than it will reduce the correlations 
in a classical state. Since we have shown that in the case when 
the reference object is lossless, entanglement combined with 
bucket detection offers no improvement for neither two-photon 
imaging nor for quantum holography, we conjecture that \textit{if 
bucket detection of one of the photons is used}, entangled 
two-photon imaging never offers any advantage over (classically) 
correlated-photon imaging.

An effect similar to ``quantum holography,'' but not to be 
confused with it, has been demonstrated in 
Refs.~\cite{Strekalov,Pittman}, where a diffraction pattern could 
be read out trough the joint probability distribution of a 
two-photon state, although the marginal probability distributions 
did not show any diffraction pattern. This is an effect due to 
entanglement and the fact that \textit{bucket detection was not 
used}. To model such an effect in the simplest possible fashion, 
consider the two-photon, four-mode, entangled state \beq 
\ket{\Psi} = \frac{1}{2}(\ket{1_1,1_{1'}}+\ket{1_1,1_{2'}}+ 
\ket{1_2,1_{1'}}-\ket{1_2,1_{2'}}) . \label{Eq: demonstration 
state} \eeq Assume that $\hone = \hat{1}_1$ and 
$h_2(1,1)=h_2(1,2)= h_2(2,1)= - h_2(2,2)= 1/\sqrt{2}$. One can 
then readily show that $p(1,1')=p(2,2')=1/2$ and 
$p(1,2')=p(2,1')=0$. That is, which one of the (two) detectors 
behind each object that will ``click'' is totally unpredictable. 
The statistics of either detector pair alone do not contain any 
information about object 2. This is so regardless if one simply 
ignores the outcome of the other detector pair, or if one uses a 
bucket detector (that, in this case, always will ``click''). Due 
to the input-state entanglement, there is, however, a perfect 
correlation between the joint detection events. The correlation 
statistics hence reveal (some) information about the object.

In summary, we have shown that although entangled two-photon 
imaging may bring out effects that cannot be mimicked by any 
classical source, the advantage with this imaging method is lost 
if ``bucket detection'' of one of the photons is employed, at 
least if the reference object is lossless (and this ideal case is 
what one should strive for, since it gives the optimal 
information about the test object). To capitalize on the unique 
properties of two-photon imaging, the detection joint probability 
distribution needs to be retained.

\begin{acknowledgments}
The authors would like to thank Professor B. E. A. Saleh for 
useful comments on this work. The work was supported by the 
Swedish Research Council (VR) and the Swedish Foundation for 
Strategic Research (SSF).
\end{acknowledgments}

\end{document}